\documentclass{amsart}

\usepackage{t1enc}
\usepackage[utf8]{inputenc}

\usepackage{latexsym,pifont}
\usepackage{epsfig}
\usepackage{rotating}

\usepackage{ifpdf}
\ifpdf
\usepackage{epstopdf}
\usepackage{hyperref}
\else
\usepackage[hypertex]{hyperref}
\fi

\usepackage{amsfonts,amsma th,amssymb,mathrsfs,extarrows,MnSymbol}
\usepackage[all]{xy}

\usepackage{tikz}
\usetikzlibrary{matrix,arrows}

\newcommand{\alxydim}[2]{\begin{aligned}\xymatrix#1{#2}\end{aligned}}

\newcommand{\brem}{\begin{Rem}}
\newcommand{\erem}{\end{Rem}\medskip}
\newcommand{\beg}{\begin{Eg}}
\newcommand{\eeg}{\end{Eg}}
\newcommand{\bedef}{\begin{Def}}
\newcommand{\exdef}{\begin{flushright}$\diamond$\end{flushright}
\end{Def}\vskip0.1cm}
\newcommand{\berop}{\begin{Prop}}
\newcommand{\eerop}{\end{Prop}}
\newcommand{\belem}{\begin{Lem}}
\newcommand{\elem}{\end{Lem}}
\newcommand{\bethe}{\begin{Thm}}
\newcommand{\ethe}{\end{Thm}}
\newcommand{\becor}{\begin{Cor}}
\newcommand{\ecor}{\end{Cor}}
\newcommand{\beroof}{\noindent\begin{proof}}
\newcommand{\eroof}{\end{proof}}
\newcommand{\becon}{\begin{Conv}}
\newcommand{\econ}{\begin{flushright}$\checkmark$\end{flushright}\end{Conv}}
\newcommand{\befact}{\begin{Fact}}
\newcommand{\efact}{\begin{flushright}$\checkmark$\end{flushright}\end{Fact}}
\newcommand{\bequest}{\begin{Quest}}
\newcommand{\equest}{\end{Quest}}
\newcommand{\brob}{\begin{Prob}}
\newcommand{\erob}{\end{Prob}}
\newcommand{\becj}{\begin{conj}}
\newcommand{\ecj}{\begin{flushright}$\boxtimes$\end{flushright}\end{conj}}

\newcommand{\barr}{\begin{array}}
\newcommand{\earr}{\end{array}}
\newcommand{\ben}{\begin{enumerate}}
\newcommand{\een}{\end{enumerate}}
\newcommand{\bit}{\begin{itemize}}
\newcommand{\eit}{\end{itemize}}

\newcommand{\qq}{\begin{eqnarray}}
\newcommand{\qqq}{\end{eqnarray}}

\newcommand{\nn}{\nonumber}

\newcommand{\ovl}[1]{\overline{#1}}
\newcommand{\unl}[1]{\underline{#1}}

\newcommand\void[1]{}

\newcommand{\gt}[1]{\mathfrak{#1}}

\def\cA{\mathcal{A}}

\def\cD{\mathcal{D}}
\def\cE{\mathcal{E}}
\def\cF{\mathcal{F}}
\def\cG{\mathcal{G}}

\def\cI{\mathcal{I}}

\def\ceL{\mathcal{L}}
\def\cM{\mathcal{M}}

\def\cO{\mathcal{O}}

\def\cS{\mathcal{S}}
\def\cT{\mathcal{T}}

\def\cW{\mathcal{W}}

\def\cZ{\mathcal{Z}}

\def\xcG{\mathscr{G}}

\def\bC{{\mathbb{C}}}

\def\bH{{\mathbb{H}}}

\def\bN{{\mathbb{N}}}

\def\bR{{\mathbb{R}}}
\def\bS{{\mathbb{S}}}
\def\bT{{\mathbb{T}}}

\def\bZ{{\mathbb{Z}}}

\def\a{\alpha}
\def\b{\beta}
\def\g{\gamma}
\def\G{\Gamma}

\def\D{\Delta}

\def\k{\kappa}

\def\la{\lambda}

\def\om{\omega}
\def\Om{\Omega}

\def\si{\sigma}
\def\Si{\Sigma}

\def\z{\zeta}

\def\agt{\gt{a}}

\def\lgt{\gt{l}}

\def\tgt{\gt{t}}

\newcommand{\sfB}{{\mathsf B}}

\newcommand{\sfd}{{\mathsf d}}

\newcommand{\sfE}{{\mathsf E}}

\newcommand{\sfi}{{\mathsf i}}

\newcommand{\sfL}{{\mathsf L}}

\newcommand{\sfN}{{\mathsf N}}

\newcommand{\sfP}{{\mathsf P}}

\newcommand{\sfT}{{\mathsf T}}

\newcommand{\sfY}{{\mathsf Y}}

\newcommand{\txA}{{\rm A}}
\newcommand{\txb}{{\rm b}}
\newcommand{\txB}{{\rm B}}

\newcommand{\txF}{{\rm F}}
\newcommand{\txg}{{\rm g}}
\newcommand{\txG}{{\rm G}}

\newcommand{\txH}{{\rm H}}

\newcommand{\txK}{{\rm K}}

\newcommand{\txm}{{\rm m}}

\def\exp{{\rm exp}}
\def\id{{\rm id}}
\newcommand{\pr}{{\rm pr}}

\def\too{\longrightarrow}
\def\ev{{\rm ev}}

\def\1morf{1{\rm -Mor}}
\def\2morf{2{\rm -Mor}}
\def\dim{{\rm dim}}

\newcommand{\sMan}{{\rm {\bf sMan}}}

\newcommand{\sLieGrp}{{\rm {\bf sLieGrp}}}

\def\Vol{{\rm Vol}}

\def\p{\partial}

\def\emb{\hookrightarrow}

\def\curv{{\rm curv}}

\def\bd1{{\boldsymbol{1}}}
\def\brd0{{\boldsymbol{0}}}

\def\Ad{{\rm Ad}}

\newcommand{\uj}{{\rm U}(1)}

\def\x{\times}
\def\ox{\otimes}

\def\lx{{\hspace{-0.04cm}\ltimes\hspace{-0.05cm}}}

\def\lact{\vartriangleright}

\newcommand{\corr}[1]{\left\langle #1 \right\rangle}

\newtheorem{Thm}{Theorem}
\newtheorem{Prop}[Thm]{Proposition}
\newtheorem{Lem}[Thm]{Lemma}
\newtheorem{conj}{Conjecture}
\newtheorem{Cor}[Thm]{Corollary}
\theoremstyle{definition}
\newtheorem{Rem}[Thm]{Remark}
\newtheorem{Def}[Thm]{Definition}
\newtheorem{Eg}[Thm]{Example}
\newtheorem{Conv}[Thm]{Convention}
\newtheorem{Fact}[Thm]{Fact}
\newtheorem{Quest}[Thm]{Question}
\newtheorem{Prob}[Thm]{Problem}

\begin{document}
\title{A Cartan tale of the orbifold superstring}

\author{Rafa\l ~R.\ ~Suszek}
\address{R.R.S.:\ Katedra Metod Matematycznych Fizyki,\ Wydzia\l ~Fizyki
Uniwersytetu Warszawskiego,\ ul.\ Pasteura 5,\ PL-02-093 Warszawa,
Poland} \email{suszek@fuw.edu.pl}

\begin{abstract}
A geometrisation scheme internal to the category of Lie supergroups is discussed for the supersymmetric de Rham cocycles on the super-Minkowski group $\,\bT\,$ which determine the standard super-$p$-brane dynamics with that target,\ and interpreted within Cartan's approach to the modelling of orbispaces of group actions by homotopy quotients.\ The ensuing higher geometric objects are shown to carry a canonical equivariant structure for the action of a discrete subgroup of $\,\bT$,\ which results in their descent to the corresponding orbifolds of $\,\bT\,$ and in the emergence of a novel class of superfield theories with defects.
\end{abstract}

\maketitle

\section{Introduction}

Among the many instantiations of the symmetry principle in physics,\ the modelling of dynamics with internal degrees of freedom (iDOFs) represented by orbispaces of group actions figures as a particularly subtle conceptually yet robust one:\ Indeed,\ while it grants us access to potentially singular configurational geometries and necessitates,\ in the case of continous symmetries,\ the incorporation of extra (gauge-field) iDOFs with their own intricate dynamics,\ it does,\ all this complexity notwithstanding,\ sit at the core of,\ {\it i.a.},\ the Standard Model of fundamental interactions,\ and so plays the r\^ole of an organising principle in our unified picture of elementary processes in a vast energy range.\ Being rooted in Cartan's idea of the homotopy quotient \cite{Cartan:1950mix},\ the modelling naturally leads to the emergence of meshes of codimension-1 \emph{spacetime defects} in the field theory with a gauge(d) symmetry,\ at which there occur field discontinuities determined by the action of the symmetry group.\ This entails further structural subtleties in the dynamics in the presence of non-tensorial couplings,\ such as,\ {\it e.g.},\ the topological Wess--Zumino (WZ) couplings in the nonlinear $\si$-model,\ employed successfully in the description of a wide range of dynamical systems -- from the Affleck--Haldane critical field theory of collective excitations of quantum spin chains all the way to classical (super)string theory.\ Here,\ Cartan's mixing construction \cite{Tu:2020} needs to be lifted to the higher-geometric (HG) and -categorial objects (a.k.a.\ $p$-gerbes and their morphisms) which represent the background $(p+2)$-form fields over the space of iDOFs,\ and so it undergoes a categorification in the form of an equivariant structure on these objects,\ whose components decorate the aforementioned gauge-symmetry defects \cite{Gawedzki:2010rn,Suszek:2012ddg}.

In the present note,\ we embed the above general discussion in the setting of the Green--Schwarz (GS) superfield theory \cite{Green:1983wt} of extended distributions of supercharge (a.k.a.\ super-$p$-branes) in the super-Minkowski target Lie supergroup $\,\bT\equiv\bR^{d,1|D_{d,1}}\,$ (with $\,D_{d,1}\,$ supercharges -- {\it cp.}\ later),\ the relevance of these superfield theories hinging firmly upon their asymptotic relation to the super-$\si$-models with curved supertargets with bodies $\,{\rm AdS}_m\x\bS^n$,\ which underly the physically useful yet mathematically still largely elusive AdS/CFT correspondence.\ As the point of departure,\ we take the geometrisation scheme developed in \cite{Suszek:2023ldu} for the non-trivial classes in the Cartan--Eilenberg (CaE) cohomology $\,{\rm CaE}^\bullet(\bT)\equiv H^\bullet(\bT)^\bT\,$ of that Lie supergroup which determine the relevant WZ couplings.\ The scheme,\ to be regarded as a generalisation of the intrinsically FDA construction of the so-called \emph{extended superspaces} due to de Azc\'arraga {\it et al.} \cite{Chryssomalakos:2000xd},\ yields distinguished Murray--Stevenson-type $p$-gerbe objects \cite{Murray:1994db,Stevenson:2001grb2} in the category $\,\sLieGrp\,$ of Lie supergroups,\ dubbed {\bf CaE super-$p$-gerbes} in \cite{Suszek:2023ldu}.\ Upon reviewing the logic of their construction,\ we identify the resultant HG objects as $p$-gerbes over the super-minkowskian base endowed with a \emph{canonical} equivariant structure for the \emph{discrete} Kosteleck\'y--Rabin supersymmetry group $\,\G_{\rm KR}\subset\bT\,$ of \cite{Kostelecky:1983qu}.\ As such,\ they are to be viewed -- in the spirit of an HG extension of Cartan's construction worked out in \cite{Gawedzki:2010rn,Suszek:2012ddg} -- as \emph{models} of $p$-gerbes over the {\bf Rabin--Crane super-orbifold} $\,\bT//\G_{\rm KR}\,$ of their base.\ The super-orbifold has \emph{compact Gra\ss mann-odd fibres} \cite{Rabin:1984rm} and so seems to be \emph{singular} as a supergeometry \cite{Rothstein:1986ax} -- this is a situation in which Cartan's idea becomes particularly relevant and useful,\ even indispensable.\ The above identification is a nontrivial extension of Rabin's original proposal \cite{Rabin:1985tv} to view $\,{\rm CaE}^\bullet(\bT)\,$ as the differential cohomology \emph{equivariant} with respect to $\,\G_{\rm KR}$,\ and therefore \cite{Tu:2020} -- as a \emph{model} of the de Rham cohomology of $\,\bT//\G_{\rm KR}$.\ Finally,\ the derivation of the canonical $\G_{\rm KR}$-equivariant structure on the super-$p$-gerbes of the GS super-$\si$-model provides us with a novel construction of a superfield theory with $\G_{\rm KR}$-jump defects,\ which is to be understood,\ along the lines of \cite{Runkel:2008gr,Suszek:2012ddg},\ as an \emph{effective model} of super-$p$-brane dynamics on $\,\bT//\G_{\rm KR}$.\ Superstring theory is thus shown to \emph{naturally} inhabit such singular supergeometries,\ which are yet to be explored.

\section{Configurational \& dynamical aspects of Cartan's mixing construction}

The subject of our interest is a lagrangean field theory with the space of iDOFs given by the space of orbits $\,M//\G\,$ of an action $\,\la:\G\x M\to M:(g,x)\mapsto\la_g(x)\equiv g\lact x\,$ of a group $\,\G\,$ on a smooth manifold $\,M$.\ If $\,\la\,$ is \emph{not} free and proper,\ there may be no smooth structure on and hence no \emph{direct} access to $\,M//\G$.\ Under such circumstances,\ we invoke Cartan's mixing construction \cite{Cartan:1950mix,Tu:2020} and \emph{model} the field theory on classes of field configurations in $\,M$.\ Thus,\ the point of departure is a field theory with $\,M\,$ as the space of iDOFs,\ {\it i.e.},\ a {\bf field bundle} $\,\pi_\cF:\cF\to\Si\,$ with typical fibre $\,M\,$ over a metric {\bf spacetime} $\,(\Si,\unl\txg)\,$ together with a functional $\,\cA_{\rm DF}\equiv\exp(\sfi\,S):\G(\cF)\to\uj$,\ termed the {\bf Dirac--Feynman amplitude} (DFA),\ whose critical points are the classical field configurations.\ The DFA is expressed in terms of an {\bf action functional} $\,S$,\ itself determined by a lagrangean density as $\,S[\phi]=\int_\Si\,\ceL(\phi,\sfT\phi)$.\ The identification of $\,\cA_{\rm DF}\,$ as the fundamental object is in keeping with Dirac's quantum-mechanical interpretation of $\,S$,\ and it paves the way towards a rigorous description of charge dynamics over topologically non-trivial spacetimes \cite{Gawedzki:1987ak},\ as in the Aharonov--Bohm experiment.\ We assume $\,\cA_{\rm DF}\,$ to be $\la$-invariant and subsequently `descend' the field theory from $\,M\,$ to $\,M//\G$.\ To this end,\ we consider {\bf Cartan's mixing diagram} \cite{Tu:2020}
\qq\label{diag:Cartmix}
\alxydim{@C=1.5cm@R=.75cm}{ \sfE\G \ar[d]_{\pi_{\sfE\G}} & \sfE\G\x M \ar[d]_{\pi_\sim} \ar[l]^{\pr_1\ } \ar[r]_{\quad\pr_2} & M \ar@{..>}[d]^{\varpi_\sim} \\ \sfB\G & \sfE\G\x_\la M \ar[l] \ar@{..>}[r]^{\pi} & M//\G}\,,
\qqq
in which $\,\pi_{\sfE\G}:\sfE\G\to\sfB\G\,$ is the universal principal $\G$-bundle over the classifying space $\,\sfB\G\,$ of $\,\G$,\ with a \emph{contractible} total space $\,\sfE\G\,$ on which $\,\G\,$ acts,\ and in which $\,\sfE\G\x_\la M\equiv(\sfE\G\x M)//\G\,$ is the quotient \emph{manifold} for the diagonal action of $\,\G$.\ The latter manifold fibres over $\,\sfB\G\,$ with typical fibre $\,M$,\ and admits a fibrewise action of the {\bf universal adjoint bundle} $\,\Ad\sfE\G\equiv\sfE\G\x_\Ad\G$,\ locally modelled on $\,\la$.\ Whenever $\,M//\G\,$ is smooth,\ with the orbit projection $\,\varpi_\sim\,$ smooth and $\G$-principal,\ it also fibres over $\,M//\G\,$ as $\,\pi:\sfE\G\x_\la M\to M//\G:[(p,x)]\mapsto\G\lact x$,\ which implies the homotopy equivalence $\,\sfE\G\x_\la M\sim_{\rm h}M//\G$.\ The equivalence underlies Cartan's identification of the {\bf universal associated bundle} $\,\sfE\G\x_\la M\,$ as a \emph{smooth model} of the potentially singular quotient geometry $\,M//\G$,\ carrying information on the latter's homotopy-invariant structures -- hence,\ {\it e.g.},\ the definition of the orbispace de Rham cohomology $\,H^\bullet(M//\G):=H^\bullet(\sfE\G\x_\la M)$,\ a.k.a.\ the {\bf $\G$-equivariant cohomology of} $\,M\,$ \cite{Tu:2020}.\ In the field-theoretic setting of interest,\ we work with avatars of Cartan's universal construction in the form of pullbacks of \eqref{diag:Cartmix} along maps $\,\Phi:\Si\to\sfB\G$,\ {\it i.e.},\ with {\bf gauge bundles} $\,\sfP_\Phi\equiv\Phi^*\sfE\G\,$ and the {\bf associated bundles} $\,\sfP_\Phi M\equiv\Phi^*(\sfE\G\x_\la M)\,$ whose global sections $\,\G(\sfP_\Phi M)\,$ acquire the status of (matter) fields of the theory with $\,\G\,$ gauged.\ These are acted upon by the {\bf gauge group} $\,\G(\Ad\sfP_\Phi)$,\ capturing \emph{vertical auto-equivalences} $\,{\rm Aut}_{{\rm {\bf Bun}}_\G(\Si)}^{\rm vert}(\sfP_\Phi)\cong\G(\Ad\sfP_\Phi)\,$ of the field-theoretic pullback.\ The relevance of these structures to the attainment of the original goal is readily appreciated when $\,M//\G\,$ is smooth:\ In this case,\ sections $\,\phi\equiv(\id_\Si,f)\in\G(\sfP_\Phi M)\,$ of $\,\sfP_\Phi M\equiv\Si\x_\Phi(\sfE\G\x_\la M)\,$ induce smooth maps $\,\pi\circ f\in[\Si,M//\G]$.\ In a generic situation,\ the picture is subtler but no less meaningful:\  Taking into account the classic bijection $\,\Psi_\la:\G(\sfP_\Phi M)\cong{\rm Hom}_\G(\sfP_\Phi,M)\,$ between matter fields and $\G$-equivariant maps $\,\sfP_\Phi\to M$,\ and using sections $\,\si_i:\cO_i\to\sfP_\Phi,\ i\in I\,$ of $\,\sfP_\Phi\,$ over a trivialising cover $\,\{\cO_i\}_{i\in I}$,\ we represent $\,\phi\,$ \emph{locally} by smooth maps $\,f_i\equiv\si_i^*\Psi_\la[\phi]:\cO_i\to M$,\ subject to identifications $\,f_i(y)=\la_{g_{ij}(y)}(f_j(y))$ over the $\,\cO_{ij}\equiv\cO_i\cap\cO_j\ni y\,$ effected by the transition mappings $\,g_{ij}:\cO_{ij}\to\G\,$ of $\,\sfP_\Phi\,$ associated with the {\bf local gauges} $\,\si_i$.\ The {\bf local matter fields} $\,f_i\,$ undergo {\bf local gauge transformations} $\,f_i(\cdot)\longmapsto\la_{\g_i(\cdot)}(f_i(\cdot))\,$ determined by presentations $\,\g_i\equiv\si_i^*\Psi_\Ad[\g]:\cO_i\to\G\,$ of gauge-group elements $\,\g\in\G(\Ad\sfP_\Phi)$.\ Thus,\ configurations of the orbispace field theory are \emph{effectively realised} as equivalence classes,\ with respect to the action of $\Si$-dependent (local) profiles in $\,\G$,\ of \emph{piecewise} smooth profiles in $\,M\,$ over domains $\,\cO_i\,$ of (arbitrary) trivialisations of $\,\sfP_\Phi M$,\ with discontinuities at arbitrarily located (within the $\,\cO_{ij}$) `domain walls' controlled by profiles $\,g_{ij}\,$ in $\,\G\,$ (which `smoothen out' upon projection to $\,M//\G$).\ In this manner,\ Cartan's construction gives rise to a field theory \emph{with topological gauge-symmetry defects},\ {\it i.e.},\ codimension-1 {\it loci} of field discontinuity which can be continuously deformed within the spacetime of the theory without affecting the value of its DFA,\ {\it cp.}\ \cite{Runkel:2008gr,Suszek:2012ddg} for details.\ It deserves to be emphasised that it is the existence of \emph{nontrivial} gauge bundles $\,\sfP_\Phi\,$ which entails the emergence of the so-called {\bf twisted sector} (with properly discontinuous field configurations) in the orbispace field theory.

Whenever $\,\G\,$ is smooth,\ the above configurational construction is not the end of the story:\ Describing the \emph{dynamics} of the iDOFs from $\,M//\G\,$ calls for the incorporation of a $\G(\Ad\sfE\G)$-equivariant connection on $\,\sfE\G\x_\la M\,$ canonically induced from a principal one on $\,\sfE\G$.\ As the symmetry group of our immediate interest is discrete,\ which we assume with regard to $\,\G\,$ henceforth,\ we do not elaborate this point,\ referring the Reader to the literature,\ {\it e.g.},\ \cite{Gawedzki:2010rn,Suszek:2012ddg}.

The construction of the orbispace field theory uses differential-geometric structures over $\,M$,\ which enter the definition of the mother DFA.\ In an attempt to `descend' the latter to $\,M//\G$,\ we may encounter two distinct situations:\ (i) the dynamics is determined by $\la$-invariant tensors on $\,M$,\ {\it e.g.},\ the standard `kinetic' term given by a metric tensor $\,\xcG\in\G(\sfT^*M\ox_{M,\bR}\sfT^*M)\,$ on the space of iDOFs with $\,{\rm Isom}(M,\xcG)\supset\G$;\ (ii) the dynamics is determined by non-tensorial couplings sourced by (higher-)geometric objects over $\,M\,$ which couple to the `states' of the field theory,\ and the DFA is invariant under automorphisms of these objects,\ including lifts of the $\,\la_g$.\ In the former case,\ the descent is straightforward:\ Any $\G$-invariant tensor is automatically $\G$-basic for $\,\G\,$ discrete,\ and so the DFA for $\,[\Si,M]$,\ \emph{as it stands},\ describes the iDOFs \emph{modulo} $\,\la$.\ In particular,\ in the previously described defect picture of Cartan's model,\ there is no need to augment the patchwise definition of the twisted sector over the $\,\cO_i\,$ with any extra data supported on the defect graph -- the discrete jumps at the defect lines do not introduce any inconsistencies,\ and the lines themselves can be deformed freely within the intersections $\,\cO_{ij}\,$ of the trivialisation frames.\ In the latter scenario,\ on the other hand,\ the situation may become substantially more involved,\ which we demonstrate on a specific example,\ with view to subsequent considerations.

We focus on the lagrangean field theory -- the so-called {\bf 2$d$ $\si$-model} -- of smooth mappings $\,[\Si,M]\,$ from a closed\footnote{The assumption of closedness of $\,\Si$,\ adopted for simplicity,\ may readily be dropped in a systematic manner \cite{Runkel:2008gr}.} two-dimensional manifold $\,\Si\in\p^{-1}\emptyset\,$ into a metric manifold $\,(M,\txg)$,\ which describes minimal embeddings of the {\bf worldsheet} $\,\Si\,$ in the {\bf target space} $\,M\,$ distorted by Lorentz-type forces induced by a torsion field $\,\txH\in Z^3_{\rm dR}(M)\,$ that couples to the fundamental charged loop (a connected component $\,\cong\bS^1\,$ of an equitemporal slice in $\,\Si$).\ Whenever $\,0\neq[\txH]\in H^3_{\rm dR}(M)$,\ and such a choice may be enforced upon us,\ {\it e.g.},\ by the requirement of conformality of the quantised theory,\ the topological WZ term in the DFA,\ which models the coupling of $\,\txH$,\ does not admit a simple tensorial definition -- instead,\ the amplitude takes the form \cite{Gawedzki:1987ak}
\qq\label{eq:sigmAmpl}\qquad\qquad
\cA_{\rm DF}\ :\ [\Si,M]\too\uj\ :\ x\longmapsto\exp\left(\sfi\int_\Si\Vol(\Si,x^*\txg)\right)\cdot{\rm Hol}_\cG\bigl(x(\Si)\bigr)\,,
\qqq
given in terms of the {\bf surface holonomy} $\,{\rm Hol}_\cG(x(\Si))\,$ of a {\bf 1-gerbe} $\,\cG\,$ of curvature $\,\txH\equiv{\rm curv}(\cG)$,\ calculated along $\,x(\Si)\subset M$.\ The 1-gerbe is a geometrisation of the \emph{integral} class $\,[\txH]$,\ described by the {\bf Murray diagram} \cite{Murray:1994db}
\qq\hspace{-.5cm}
\cG\quad :\ \alxydim{@C=1.25cm@R=.75cm}{ \unl\D{}^{(3)}\mu_L=\id \ar@{->>}[d] & \mu_L:\unl\D{}^{(2)}L\cong I_0 \ar@{->>}[d] & L,\cA_L \ar@{->>}[d]_{\pi_L} & & \\ \sfY^{[4]}M \ar@{->>}[r]^{\unl d{}^{(3)}_\cdot} & \sfY^{[3]}M \ar@{->>}[r]^{\unl d{}^{(2)}_\cdot} & \sfY^{[2]}M \ar@{->>}[r]^{\unl d{}^{(1)}_\cdot} & \sfY M,\txB \ar@{->>}[r]^{\pi_{\sfY M}} & M,\txH}\nn\\ \label{diag:Murray}
\qqq
over Segal's nerve $\,\sfN_\bullet{\rm Pair}_M(\sfY M)\equiv\sfY^{[\bullet+1]}M\,$ of the $M$-fibred pair groupoid $\,{\rm Pair}_M(\sfY M)\,$ (with the arrow manifold $\,\sfY^{[2]}M\equiv\{(y_1,y_2)\in\sfY M^{\x 2}\ |\ \pi_{\sfY M}(y_1)=\pi_{\sfY M}(y_2)\}$) of a surjective submersion $\,\sfY M\,$ over $\,M\,$ which supports a smooth primitive $\,\txB\,$ of $\,\pi_{\sfY M}^*\txH=\sfd\txB$.\ The $\,\unl d{}^{(n)}_k,\ k\in\ovl{0,n},\ n\in\bN^\x\,$ are the face maps of $\,\sfN_\bullet{\rm Pair}_M(\sfY M)$,\ inherited from the standard pair groupoid $\,{\rm Pair}(\sfY M)\,$ through restriction and used to define pullback operators $\,\unl\D{}^{(n)}:=\sum_{k=0}^n(-1)^k\,\unl d{}^{(n)\,*}_k$,\ and we have additional structure:\ a principal $\bC^\x$-bundle $\,L\,$ with a principal connection form $\,\cA_L\in\Om^1(L)\,$ of curvature $\,\unl\D{}^{(1)}\txB$,\ and a connection-preserving isomorphism $\,\mu_L:\pr_{1,2}^*L\ox\pr_{2,3}^*L\cong\pr_{1,3}^*L\,$ ($I_0\,$ is the trivial bundle with null connection) which equips $\,L\,$ with a (fibrewise) groupoid structure.\ 1-gerbes over a given base $\,M\,$ form a bicategory $\,\gt{Grb}_\nabla(M)\,$ with a tensor product $\,\ox$,\ and we have pullback 2-functors for smooth base maps \cite{Waldorf:2007mm}.\ In fact,\ 1-gerbes belong to a hierarchy of geometrisations of classes in $\,H^{n+2}(M,\bZ),\ n\in\bN$,\ termed {\bf $n$-gerbes} (with principal $\bC^\x$-bundles with connection as $0$-gerbes) and described by respective Murray diagrams which share a structural property,\ implicit in \eqref{diag:Murray}:\ An $(n+1)$-gerbe over $\,M\,$ is represented by a smooth primitive $\,\b\in\Om^{n+2}(\sfY M)\,$ of its curvature on a surjective submersion $\,\sfY M\to M$,\ alongside an $n$-gerbe of curvature $\,\unl\D{}^{(1)}\b\,$ over $\,\sfY^{[2]}M\,$ and a collection of $k$-cells ($0<k\leq n+1$) in the (weak) $(n+1)$-categories $\,\gt{Grb}^{(n)}_\nabla(\sfY^{[k+2]}M)\,$ of $n$-gerbes over the respective components of $\,\sfN_\bullet{\rm Pair}_M(\sfY M)$,\ the last of which -- an $(n+1)$-cell in $\,\gt{Grb}^{(n)}_\nabla(\sfY^{[n+3]}M)\,$ -- is a connection-preserving principal $\bC^\x$-bundle isomorphism $\,\mu$,\ subject to a coherence constraint $\,\unl\D{}^{(n+2)}\mu=\id\,$ over $\,\sfY^{[n+4]}M\,$ ({\it cp.},\ {\it e.g.},\ \cite{Stevenson:2001grb2}).

The holonomy can now be understood as the image $\,\imath([x^*\cG])\equiv{\rm Hol}_\cG(x(\Si))\,$ of the pullback element $\,[x^*\cG]\,$ in the group $\,\cW^3(\Si;\curv=0)\,$ of isoclasses of flat 1-gerbes over $\,\Si\,$ (with the binary operation $\,\ox\,$ on representatives) under the canonical isomorphism $\,\imath:\cW^3(\Si;0)\cong H^2(\Si,\uj)\equiv\uj$.\ For the cohomologically trivial $\,\txH=\sfd\txb\,$ with $\,\txb\in\Om^2(M)$,\ we obtain the {\bf trivial gerbe} $\,\cI_\txb\,$ with $\,(\sfY M,\pi_{\sfY M},\txB,L,\pi_L,\cA_L,\mu_L\equiv\bd1)\equiv(M,\id_M,\txb,M\x\bC^\x,\pr_1,\pr_2^*\theta_{\rm L},(x,z_1)\ox(x,z_2)\longmapsto(x,z_1\cdot z_2))$,\ and the expected formula $\,{\rm Hol}_{\cI_\txb}(x(\Si))=\exp(\sfi\int_\Si x^*\txb)$.\ It is to be emphasised that the fully fledged bicategorial structure is requisite for a consistent modelling -- in terms of \emph{decorated} surface holonomies -- of defects with self-intersections in the spacetime $\,\Si\,$ of the theory whose `bulk' dynamics is determined by the 1-gerbe:\ In this case,\ the target space becomes stratified as $\,M\sqcup Q\sqcup\bigsqcup_{n\geq 3}\,T_n$,\ reflecting the decomposition of $\,\Si$,\ effected by the embedded defect graph $\,\xcG$,\ into (closed) 2$d$ domains $\,\Si_i$,\ 1$d$ domain walls (or defect lines) $\,\ell_{i,j}=\Si_i\cap\Si_j\subset\xcG\,$ and their junctions $\,v_{i_1,i_2,\ldots,i_n}\,$ of valence $\,n\in\bN_{\geq 3}$.\ Each connected bulk (sub)stratum $\,M_i\subset M\,$ comes with its own metric and a 1-gerbe, to be pulled back to $\,\Si_i$;\ each defect-line (sub)stratum $\,Q_{i,j}\subset Q\,$ comes with a 1-cell of $\,\gt{Grb}_\nabla(Q_{i,j})$,\ to be pulled back to $\,\ell_{i,j}$;\ and,\ finally,\ each junction (sub)stratum $\,T_{i_1,i_2,\ldots,i_n}\subset T_n\,$ comes with a 2-cell of $\,\gt{Grb}_\nabla(T_{i_1,i_2,\ldots,i_n})$,\ to be pulled back to $\,v_{i_1,i_2,\ldots,i_n}$,\ {\it cp.}\ \cite{Runkel:2008gr} for details.\ The fundamental benefit of working with 1-gerbes is that they canonically determine prequantisation of the 2$d$ field theory through cohomological transgression $\,\bH^2(M,\cD(2)^\bullet)\ni[\cG]\longmapsto[\ceL_\cG]\in\bH^1(\sfL M,\cD(1)^\bullet)$,\ {\it i.e.},\ a canonical assignment to (the isoclass of) $\,\cG\,$ of (the isoclass of) a line bundle over the (single-loop) configuration space $\,\sfL M\equiv[\bS^1,M]\,$ with connection of curvature $\,\int_{\bS^1}\,\ev^*\txH$,\ as discovered in the seminal work \cite{Gawedzki:1987ak}.\ The latter fact justifies adherence,\ subsequent to this discovery,\ to the principle of `gerbification' ({\it i.e.},\ construction of a bicategorial lift) applied in the analysis of all properties and structures of the \emph{classical} 2$d$ field theory which are anticipated to survive in the quantum r\'egime,\ and explains the tremendous success of the thus based HG approach to its study,\ pioneered and developed by Gaw\c{e}dzki.

Among the bicategorial lifts,\ we find that of global symmetries of the $\si$-model:\ Given $\,\G\,$ as above,\ the lift consists of a family $\,\{\Phi_g:\cG\cong\la_g^*\cG\}_{g\in\G}\,$ of invertible 1-cells of $\,\gt{Grb}_\nabla(M)$,\ which transgress to automorphisms of $\,\ceL_\cG$.\ The data $\,\{\Phi_g\}_{g\in\G}\,$ are the point of departure for the descent $\,M\searrow M//\G\,$ in the HG setting,\ as governed by the equivalence $\,\gt{Grb}_\nabla(M//\G)\cong\gt{Grb}_\nabla(M)^{\G{\rm -equiv}}\,$ -- a specialisation of the more general one from \cite{Gawedzki:2010rn} to the case of $\,\G\,$ discrete -- between the bicategory of 1-gerbes over the quotient manifold $\,M//\G\,$ (whenever it exists) and that of 1-gerbes over $\,M\,$ equipped with a {\bf $\G$-equivariant structure},\ {\it i.e.},\ simplicial 1-gerbes over the nerve $\,\sfN_\bullet(\G\lx_\la M)\equiv\G^{\x\bullet}\x M\,$ of the action groupoid $\,\G\lx_\la M$,\ represented by diagrams
\qq\nn
\alxydim{@C=1.25cm@R=.75cm}{ \D^{(3)}\upsilon=\id \ar@{->>}[d] & \upsilon:\D^{(2)}\Upsilon\cong I_0 \ar@{->>}[d] & \Upsilon:\D^{(1)}\cG\cong\cI_0 \ar@{->>}[d] & & \\ \G^{\x 3}\x M \ar@{->>}[r]^{d^{(3)}_\cdot} & \G^{\x 2}\x M \ar@{->>}[r]^{d^{(2)}_\cdot} & \G\x M \ar@{->>}[r]^{d^{(1)}_\cdot} & M \ar@{..>>}[r]^{\varpi_\sim\quad} & M//\G}\,,
\qqq
in which the $\,d^{(n)}_k,\ k\in\ovl{0,n},\ n\in\bN^\x\,$ are the face maps of $\,\sfN_\bullet(\G\lx_\la M)$,\ defining the $\,\D{}^{(n)}:=\sum_{k=0}^n(-1)^k\,d^{(n)\,*}_k$,\ and we have additional structure:\ a 1-cell $\,\Upsilon\,$ in $\,\gt{Grb}_\nabla(\G\x M)\,$ which establishes equivalence between $\,\la^*\cG\,$ and $\,\pr_2^*\cG$,\ and a 2-cell $\,\upsilon\,$ in $\,\gt{Grb}_\nabla(\G^{\x 2}\x M)\,$ which renders the bicategorial lift $\,(\cG,\Upsilon)\,$ of $\,\la\,$ homomorphic and associative.\ In the light of the above bicategorial correspondence,\ existence of the $\G$-equivariant structure ensures descent of the pullback 1-gerbe $\,\pr_2^*\cG\,$ from over $\,\sfE\G\x M\,$ to $\,\sfE\G\x_\la M$,\ and so also after the field-theoretic pullback $\,\Phi$,\ whereby there arises a 1-gerbe $\,\unl\cG\,$ over the field bundle $\,\sfP_\Phi M\,$ such that $\,\pi_\sim^*\unl\cG\cong\pr_2^*\cG\,$ \cite{Gawedzki:2010rn}.\ The r\^ole of $\,(\cG,\Upsilon,\upsilon)\,$ becomes more intuitively clear in the local picture with gauge-symmetry defects -- indeed,\ the triple is precisely what is needed to consistently define the aforementioned decorated surface holonomy in the presence of defect lines $\,\ell_{i,j}\ni y\,$ at which the bulk embedding field $\,x_i:\cO_i\to M\,$ jumps as $\,x_i(y)=\la_{g_{ij}(y)}(x_j(y))\,$ by the (locally) constant transition mappings $\,g_{ij}:\cO_{ij}\to\G$:\ With the dynamics in the domains $\,\Si_i\,$ determined by the $\,x_i^*\cG$,\ we endow the defect lines $\,\ell_{i,j}\,$ with the respective 1-cells $\,(g_{ij},x_j)^*\Upsilon$,\ and the \emph{elementary} (valence-3) junctions $\,v_{i,j,k}\,$ with the 2-cells $\,(g_{ij},g_{jk},x_k)^*\upsilon$.\ The associator condition $\,\D^{(3)}\upsilon=\id$,\ satisfied by $\,\upsilon\,$ over $\,\G^{\x 3}\x M$,\ ensures that 2-cells required for junctions of higher valence can be consistently induced from the elementary ones through (vertical) 2-cell composition dictated by a limiting procedure in which the junction of valence $\geq 4$ is recovered by contracting all \emph{internal} edges of its \emph{arbitrary} binary-tree resolution,\ at no cost in the value of the DFA (owing to the topological nature of the defects),\ {\it cp.}\ \cite{Runkel:2008gr} for details and \cite{Suszek:2022lpf} for a general treatment of simplicial $\si$-models.\ In summary,\ we see that the data $\,(\cG,\Upsilon,\upsilon)\,$ \emph{effectively determine} an orbispace $\si$-model with target $\,M//\G$.

\section{The super-Minkowskian super-$p$-gerbes:\ construction \& supersymmetry}\label{sec:super-p-gerbes}

We now come to the main point in our story,\ at which we embed the previous constructions in the $\bZ/2\bZ$-graded geometric category,\ and add supersymmetry as the fundamental symmetry of the ensuing superfield theory.\ More specifically,\ let $\,\Si\in\p^{-1}\emptyset\,$ be a $(p+1)$-dimensional smooth manifold and fix a supermanifold $\,\cM=(|\cM|,\cO_\cM)\,$ of superdimension $\,(m|n)$,\ with a {\bf body} (manifold) $\,|\cM|\,$ and a structure sheaf $\,\cO_\cM$,\ locally modelled on $\,(\bR^{\x m},C^\infty(\cdot)\ox B_n)\equiv\bR^{m|n}\,$ for the rank-$n$ Gra\ss mann algebra $\,B_n\equiv\bigwedge^\bullet\bR^{\x n}$.\ Assume further the existence of a Lie supergroup $\,\txG=(|\txG|,\cO_\txG)\,$ (a group object in the category $\,\sMan\,$ of supermanifolds) acting on $\,\cM\,$ as $\,\la:\txG\x\cM\to\cM$.\ The supermanifold $\,\cM\,$ should come with an even rank-2 symmetric tensor $\,\txg\in\G(\cT^*\cM\ox\cT^*\cM)_{(0)}\,$ and an even integral de Rham $(p+2)$-cocycle $\,\txH_p$,\ both $\la$-invariant.\ Given the Gra\ss mann-even nature of $\,\Si$,\ probing the {\bf soul} of the supertarget $\,\cM\,$ ({\it i.e.},\ the nilpotent component of $\,\cO_\cM$) calls for an inner-${\rm Hom}$ functorial structure of the space of fields,\ which we take,\ after \cite{Freed:1999},\ in the form of the \emph{generalised} {\bf mapping supermanifold} $\,[\Si,\cM]\equiv{\rm Hom}_\sMan(\Si\x\cdot,\cM)$,\ to be evaluated on a family $\,\{\bR^{0|N}\}_{N\in\bN^\x}\,$ of {\bf superpoints},\ nested as $\,\bR^{0|N_1}\emb\bR^{0|N_2}\,$ for $\,N_1<N_2$.\ Thus,\ we may think of the superfield theory as (the direct limit of) a family of `ordinary' field theories.\ With these,\ we define a dynamics which generalises \eqref{eq:sigmAmpl}.

An obvious choice of the target,\ motivated by the postulate of irreducibility,\ is a homogeneous space $\,\cM\equiv\txG/\txK\,$ of the {\bf supersymmetry group} $\,\txG\,$ relative to a closed subgroup $\,\txK\subset|\txG|\,$ of its body,\ which -- for $\,({\rm sLie}\,\txG,{\rm Lie}\,\txK)\,$ reductive -- allows us to employ the $\txK$-basic component of the left-invariant (LI) Cartan calculus on $\,\txG\,$ in the construction of the DFA for $\,\txG/\txK$,\ as realised by local sections of $\,\txG\to\txG/\txK$.\ This is the logic underlying,\ {\it i.a.},\ the definition of the important class of super-$\si$-models with targets $\,{\rm s}({\rm AdS}_m\x\bS^n)$,\ central to the formulation of the AdS/CFT correspondence.\ In what follows,\ we focus on the simplest homogeneous geometry -- the {\bf super-Minkowski space} $\,{\rm sISO}(d,1|D_{d,1})/{\rm Spin}(d,1)\equiv\bT$,\ with its global generators of $\,\cO_\bT$:\ the odd $\,\theta^\a,\ \a\in\ovl{1,D_{d,1}}\,$ and the even $\,x^a,\ a\in\ovl{0,d}$,\ and the Lie-supergroup structure.\ Its binary operation $\,\txm_\bT\,$ has the sheaf component $\,\txm_\bT^*(\theta^\a,x^a)=(\theta^\a\ox\bd1+\bd1\ox\theta^\a,x^a\ox\bd1+\bd1\ox x^a-\frac{1}{2}\ovl\G{}^a_{\a\b}\,\theta^\a\ox\theta^\b)$,\ where the $\,\ovl\G{}^a\equiv C\G^a$,\ assumed \emph{symmetric},\ are products of the generators $\,\G^a\,$ of the Clifford algebra $\,{\rm Cliff}(\bR^{d,1})\equiv\corr{\bd1}\oplus\bigoplus_{k=1}^{d+1}\bigoplus_{a_1<a_2<\ldots<a_k=0}^d\corr{\G^{a_1 a_2\ldots a_k}\equiv\G^{a_1}\G^{a_2}\cdots\G^{a_k}}\,$ of the Minkowski space $\,\bR^{d,1}\equiv(\bR^{\x d+1},\eta)\,$ with the skew charge-conjugation matrix $\,C$,\ all in a Majorana-spinor representation $\,S_{d,1}\,$ of $\,\dim\,S_{d,1}=D_{d,1}$.\ The corresponding Lie superalgebra $\,\tgt\equiv{\rm sLie}\,\bT=\bigoplus_{\a=1}^{D_{d,1}}\corr{Q_\a}\oplus\bigoplus_{a=0}^d\corr{P_a}\,$ has the well-known structure equations:\ $\,\{Q_\a,Q_\b\}_\tgt=\ovl\G{}^a_{\a\b}\,P_a\,$ and $\,[Q_\a,P_a]_\tgt=0=[P_a,P_b]_\tgt$.\ The relevant algebra of LI differential forms $\,\Om^\bullet(\bT)^\bT=\corr{\si^\a,e^a}\,$ has generators $\,\si^\a(\theta,x)=\sfd\theta^\a\,$ and $\,e^a(\theta,x)=\sfd x^a+\frac{1}{2}\,\theta^\a\,\ovl\G{}^a_{\a\b}\,\sfd\theta^\b\equiv\eta^{-1\,ab}\,e_b(\theta,x)$.\ Upon assuming the $\,\ovl\G{}^{a_1 a_2\ldots a_p}\equiv C\G^{a_1 a_2\ldots a_p}\,$ \emph{symmetric} and imposing the {\bf Fierz identities} $\,\eta_{ab}\,\ovl\G{}^a_{(\a\b}\,\ovl\G{}^{ba_1 a_2\ldots a_{p-1}}_{\g\delta)}=0\,$ in $\,S_{d,1}$,\ which restricts the spectrum of admissible pairs $\,(d,p)\,$ \cite{Chryssomalakos:2000xd},\ we establish the crucial property:\ $\,H^\bullet(\bT)=0\neq{\rm CaE}^\bullet(\bT)\,$ of the CaE cohomology $\,{\rm CaE}^\bullet(\bT)\ni[\ovl\G{}^{a_1 a_2\ldots a_p}_{\a\b}\,\si^\a\wedge\si^\b\wedge e_{a_1}\wedge e_{a_2}\wedge \cdots\wedge e_{a_p}\equiv\chi_p]\neq 0\,$ of $\,\bT\,$ \cite{Chryssomalakos:2000xd}.\ The property attests to the existence of an intricate topology lurking beneath the plain supergeometry $\,\bT$,\ which we elucidate below,\ following the extensive study \cite{Suszek:2023ldu}.

There is a canonical choice of the (quasi-)metric on $\,\bT$,\ given by the LI lift $\,\txg=\eta_{ab}\,e^a\ox e^b\,$ of $\,\eta$,\ and the requirement of restoration of supersymmetry in the classical vacuum of the superfield theory for $\,(\txg,\txH_p)\,$ through $\k$-symmetry projection fixes the remaining choice of the $(p+2)$-cocycle $\,\txH_p\,$ for the WZ term \cite{Suszek:2020xcu}:\ In all admissible cases $\,p<10$,\ we find $\,\txH_p=q_p\,\chi_p\,$ for some $\,q_p\neq 0$,\ and so while we might write the resultant {\bf GS super-$\si$-model} in a \emph{quasi}-supersymmetric form \eqref{eq:sigmAmpl} with a \emph{trivial} WZ term given by the pullback of a de Rham primitive of $\,\txH_p$,\ the constitutive nature of $\bT$-invariance forces us to face the problem of `geometrising' $\,[\txH_p]$,\ \emph{conceptually identical} with the one encountered in the un-graded setting for the ordinary differential cohomology.\ Technically,\ the solution to it proposed in \cite{Suszek:2023ldu} generalises the construction of the so-called extended superspaces in \cite{Chryssomalakos:2000xd} and,\ as such,\ hinges on the classic bijection between the 2nd cohomology group $\,H^2({\rm sLie}\,\txG,\agt)\,$ of the tangent Lie superalgebra $\,{\rm sLie}\,\txG\,$ of a Lie supergroup $\,\txG\,$ with values in its trivial (super)commutative module $\,\agt\,$ and the set of equivalence classes of (super)central extensions of $\,{\rm sLie}\,\txG\,$ through $\,\agt$,\ the latter being captured by short exact sequences $\,\brd0\to\agt\to\widehat{{\rm sLie}\,\txG}\to{\rm sLie}\,\txG\to\brd0\,$ of Lie superalgebras.\ For $\,\agt\,$ of (super)dimension 1,\ the former is the 2nd Chevalley--Eilenberg cohomology group $\,{\rm CE}^2({\rm sLie}\,\txG)\,$ of $\,{\rm sLie}\,\txG$,\ which is canonically isomorphic with $\,{\rm CaE}^2(\txG)$,\ and so we arrive at an algorithm of \emph{sequential geometrisation} \cite{Suszek:2023ldu}:
\ben
\item[(1)] \emph{Trivialisation of} $\,\txH_p$:\ Identification of a CaE-non-exact rank-2 $\wedge$-factor $\,\txF\,$ in $\,\txH_p$,\ and subsequent integration of the corresponding (super)central extension of $\,{\rm sLie}\,\txG\,$ to a Lie-supergroup extension $\,\widehat\pi:\widehat\txG\twoheadrightarrow\txG$,\ whereby partial trivialisation of $\,\widehat\pi{}^*\chi\,$ occurs in $\,{\rm CaE}^\bullet(\widehat\txG)$.\ This step is repeated,\ with the replacement $\,(\txH_p,\txG)\mapsto(\widehat\pi{}^*\chi,\widehat\txG)$,\ until complete trivialisation is attained over an extension $\,\pi_{\sfY\txG}:\sfY\txG\twoheadrightarrow\txG$,\ {\it i.e.},\ $\,\pi_{\sfY\txG}^*\txH_p=\sfd\txB\,$ for some $\,\txB\in\Om^{p+1}(\sfY\txG)^{\sfY\txG}$.
\item[(2)] \emph{`Murraification' of the pair} $\,(\sfY\txG,\txB)$:\ Reconstruction of the full Murray diagram of the $p$-gerbe over $\,\sfN_\bullet{\rm Pair}_\txG(\sfY\txG)\,$ through repetitive application of the stepwise trivialisation scheme from point (1) to each nontrivial \emph{structural} CaE $k$-cocycle ($2\leq k\leq p+1$) encountered along the way ({\it i.e.},\ defining a sub-$(k-2)$-gerbe,\ just as in the un-graded setting).
\een
The procedure delineated above may -- {\it a priori} -- fail at some elementary ($H^2$-)stage due to nonintegrability of a Lie-superalgebra extension.\ As it happens,\ in the study,\ reported in \cite{Suszek:2023ldu},\ of the GS $(p+2)$-cocycles $\,\txH_p\,$ for $\,p\in\{0,1,2\}\,$ determined by the requirement of vacuum supersymmetry one does \emph{not} encounter such obstructions.\ Thus,\ in all these physically distinguished cases,\ we obtain a {\bf $p$-gerbe object $\,\cG_p\,$ in $\,\sLieGrp$},\ described by a Murray diagram with a Lie supergroup at each node,\ all arrows representing Lie-supergroup epimorphisms,\ and all superdifferential forms LI.\ Such higher geometric objects were dubbed {\bf CaE super-$p$-gerbes} in \cite{Suszek:2023ldu}.\ We close this section with an explicit example of the above geometrisation mechanism -- that of the CaE super-1-gerbe $\,\cG_1\,$ -- which paves the way to a topological interpretation of the advocated geometrisation scheme and a novel superfield theory.

Thus,\ assume the Fierz identity for $\,p=1\,$ and consider the CaE 3-cocycle $\,\txH_1\equiv\si^\a\wedge\txF_\a$,\ written in terms of the \emph{odd} 2-cocycles $\,\txF_\a=\eta_{ab}\,\ovl\G{}^a_{\a\b}\,\si^\b\wedge e^b$.\ Their CE counterparts $\,\om_\a\,$ give a supercentral extension $\,(\sfY\tgt\equiv\tgt\oplus\bigoplus_{\a=1}^{D_{d,1}}\corr{\cZ^\a},[\cdot,\cdot\}_{\sfY\tgt}\equiv([\cdot,\cdot\}_\tgt+\cZ^\a\ox\om_\a)\circ(\pr_1\x\pr_1)\to(\tgt,[\cdot,\cdot\}_\tgt)$,\ known as the Green superalgebra.\ It integrates to a Lie-supergroup extension $\,\pi_{\sfY\bT}\equiv\pr_1:\sfY\bT\equiv\bT\x\bR^{0|D_{d,1}}\to\bT\,$ with a binary operation $\,\sfY\txm\,$ fixed by the requirement of left-invariance of the 1-forms $\,\z_\a=\pi_{\sfY\bT}^*\om_\a+\pr_2^*\theta_\a\,$ (dual to the $\,\cZ^\a$),\ written in terms of the MC 1-forms $\,\theta_\a\equiv\sfd\xi_\a\,$ on $\,\bR^{0|D_{d,1}}\,$ (with global coordinates $\,\xi_\a$).\ On $\,\sfY\bT$,\ we find the trivialisation $\,\pi_{\sfY\bT}^*\txH_1=\sfd\txB_1\,$ with $\,\txB_1=-\pi_{\sfY\bT}^*\si^\a\wedge\z_\a$.\ The latter has $\,\txF^{(1)}\equiv\unl\D{}^{(1)}\txB_1=\pi_{\sfY\bT}^*\si^\a\wedge\sfd(\xi^{(2)}_\a-\xi^{(1)}_\a)\,$ such that $\,0\neq[\txF^{(1)}]\in{\rm CaE}^2(\sfY^{[2]}\bT)$,\ and so its CE counterpart $\,\om\,$ defines a rank-1 central extension $\,(\lgt\equiv\sfY^{[2]}\tgt\oplus\corr{Z},[\cdot,\cdot\}_\lgt\equiv([\cdot,\cdot\}_{\sfY^{[2]}\tgt}+Z\ox\om)\circ(\pr_1\x\pr_1)\to(\sfY^{[2]}\tgt,[\cdot,\cdot\}_{\sfY^{[2]}\tgt})$,\ which integrates to $\,\pi_L\equiv\pr_1:L\equiv\sfY^{[2]}\bT\x\bC^\x\to\sfY^{[2]}\bT$.\ The binary operation $\,L\txm\,$ on $\,L\,$ follows from left-invariance of $\,\cA_L=\pr_1^*\txA+\pr_2^*\theta_{\bC^\x}\,$ (the dual of $\,Z$),\ with $\,\txA(\theta,x,\xi^{(1)},\xi^{(2)})=\theta^\a\,\sfd(\xi^{(2)}_\a-\xi^{(1)}_\a)\,$ and the standard MC 1-form $\,\theta_{\bC^\x}\,$ on $\,\bC^\x$.\ The LI 1-form provides a trivialisation $\,\pi_L^*\txF^{(1)}=\sfd\cA_L$,\ and we check $\,\unl\D{}^{(2)}\txA=0$,\ whence also the groupoid structure $\,\mu_L=\bd1$,\ readily verified to be a Lie-supergroup isomorphism.\ Altogether,\ we obtain a CaE super-1-gerbe $\,(\sfY\bT,\pi_{\sfY\bT},\txB_1,L,\pi_L,\cA_L,\mu_L)\,$ as in \eqref{diag:Murray}.

The construction of $\,\cG_1\,$ for the superstring,\ and analogous constructions of the super-$0$-gerbe for the superparticle,\ and of the super-2-gerbe for the $M$-theory supermembrane form the basis of an in-depth HG study of $\k$-symmetry \cite{Suszek:2020xcu} and of a systematic reconstruction of maximally supersymmetric super-minkowskian defects \cite{Suszek:2022lpf} along the lines of \cite{Runkel:2008gr},\ {\it i.e.},\ using a supersymmetric multiplicative structure on the super-$p$-gerbe.\ We here,\ instead,\ pursue the conceptual question:\ \emph{Can we ascribe to the above geometrisation the meaning of a resolution of a nontrivial topology,\ and thereby establish a deeper correspondence with Murray's construction?}

\section{Categorification of SUSY \& descent to the Rabin--Crane super-orbifold}

An HG object associated with an integral class in the de Rham cohomology resolves the homology dual of that class.\ This simple observation puts flesh on the bones of the previous question,\ and immediately suggests a negative answer -- indeed,\ $\,\bR^{d,1|D_{d,1}}\,$ has \emph{no} nontrivial topology.\ And yet\ldots A more elementary question in the same vein was asked by Rabin in \cite{Rabin:1985tv},\ subsequent to the work \cite{Kostelecky:1983qu} on lattice supersymmetric field theory in which \emph{discrete} subgroups of $\,\bT\,$ had come up naturally,\ to wit:\ \emph{Is there a discrete subgroup $\,\G\subset\bT\,$ with the property $\,\Om^\bullet(\bT)^\G\equiv\Om^\bullet(\bT)^\bT$?}\ Clearly,\ an affirmative answer to this question would imply,\ by Cartan's logic,\ that we can think of $\,{\rm CaE}^\bullet(\bT)\,$ as a model of $\,H^\bullet(\bT//\G)$.\ In the remainder of this note,\ we review Rabin's original answer,\ and examine our scheme of geometrisation of $\,{\rm CaE}^\bullet(\bT)\,$ from this newly acquired angle.

Let us consider a nested family of sets $\,{\rm Yon}_\bT(\bR^{0|L_1})\subset{\rm Yon}_\bT(\bR^{0|L_2}),\ L_1<L_2\,$ of superpoints in $\,\bT$,\ in the image of the Yoneda embedding $\,{\rm Yon}_\bT(\cdot)\equiv{\rm Hom}_\sMan(\cdot,\bT)\,$ of $\,\bT\,$ in the category of presheaves on $\,\sMan\,$ (a.k.a.\ generalised supermanifolds).\ It deserves to be emphasised that these are precisely the sets probed by the GS super-$\si$-model in Freed's approach.\ Technically,\ the restriction to $\,{\rm Yon}_\bT(\bR^{0|L})\,$ is effected through an explicit realisation of $\,\cO_\bT\,$ in a fixed rank-$L$ Gra\ss mann algebra $\,B_L\,$ with generators $\,\b_i,\ i\in\ovl{1,L}$,\ so that $\,\theta^\a\in B_{L\,(1)}\,$ and $\,x^a\in B_{L\,(0)}\,$ and we obtain the model $\,\bT_L\equiv B_{L\,(0)}^{\x d+1}\x B_{L\,(1)}^{\x D_{d,1}}\,$ of $\,{\rm Yon}_\bT(\bR^{0|L})$.\ We may then define the {\bf Kosteleck\'y--Rabin group at level $L$} as the subset $\,\G_{{\rm KR}\,(L)}:=$\linebreak $\corr{\bZ\,\b_{i_1}\b_{i_2}\cdots\b_{i_k}\,|\,1\leq i_1<i_2<\ldots<i_k\leq L,\ k\in\ovl{1,L}}\subset\bT_L\,$ generated multiplicatively over $\,\bZ\,$ by the basis of $\,B_L$,\ only to find the desired identity $\,\Om^\bullet(\bT_L)^{\bT_L}\equiv\Om^\bullet(\bT_L)^{\G_{{\rm KR}\,(L)}}\,$ \cite{Rabin:1985tv},\ {\it cp.}\ \cite{Suszek:2023ldu} for a proof.\ The family $\,\{\G_{{\rm KR}\,(L)}\}_{L\in\bN}\,$ inherits a nesting from $\,\{\bT_L\}_{L\in\bN}$,\ and so we may pass,\ after \cite{Rabin:1985tv},\ to the direct limit $\,\G_{\rm KR}\equiv\varinjlim\G_{{\rm KR}\,(L)}$.\ At finite level,\ $\,\Om^\bullet(\bT_L)^{\bT_L}\,$ models the exterior algebra of $\,\bT_L//\G_{{\rm KR}\,(L)}$.\ An explicit construction of such an orbifold was given in \cite{Rabin:1984rm},\ and demonstrates its geometric intricacy.\ In the direct limit,\ there arises the {\bf Rabin--Crane super-orbifold} $\,\bT//\G_{\rm KR}\equiv\varinjlim\bT_L//\G_{{\rm KR}\,(L)}$,\ and we arrive at the anticipated interpretation of $\,{\rm CaE}^\bullet(\bT)\,$ as a model of $\,H^\bullet(\bT//\G_{\rm KR})\,$ \cite{Rabin:1985tv}.\ At this point,\ it becomes natural to expect that the geometrisations $\,\cG_p\,$ of the GS classes in $\,{\rm CaE}^\bullet(\bT)\,$ from Sec.\,\ref{sec:super-p-gerbes} are particular models of $p$-gerbes over $\,\bT//\G_{\rm KR}$.

The path towards verification of the latter expectation leads through categorification of the action $\,\la\equiv\txm_\bT\,$ whose cotangent lift enters the definition of $\,{\rm CaE}^\bullet(\bT)$.\ As before,\ we focus on the case $\,p=1$,\ and adopt the $\cS$-point picture for the sake of simplicity.\ The categorification now assumes the form of a family $\,\Phi_t:\cG_1\cong\la_t^*\cG_1,\ t\equiv(\theta,x)\in\bT\,$ of 1-cells of $\,\gt{Grb}_\nabla(\bT)$.\ Taking into account the definition of pullback as a universal object,\ we may next exploit the existence of the natural lifts $\,\sfY\txm_{(t,0)}\,$ and $\,L\txm_{(t,0,0,1)}\,$ of the $\,\la_t\,$ to $\,\sfY\bT\,$ and $\,L$,\ respectively,\ in conjunction with the left-invariance of $\,\txB_1\,$ and $\,\cA_L$,\ to choose $\,(\sfY\bT,\txB_1)\,$ as the surjective submersion of the pullback 1-gerbe and $\,(L,\cA_L)\,$ as its principal $\bC^\x$-bundle,\ and thus to \emph{canonically identify} $\,\la_t^*\cG_1\,$ with $\,\cG_1\,$ \cite{Suszek:2023ldu}.\ This yields the 1-cells $\,\Phi_t\equiv\id_{\cG_1}$,\ determined \emph{fully} by the bundle $\,L\,$ and the trivial groupoid structure $\,\mu_L\equiv\bd1\,$ of $\,\cG_1\,$\cite{Waldorf:2007mm}.\ Such very special form of the categorification of supersymmetry furnished by the CaE super-1-gerbe (and the other $\,\cG_p\,$ alike),\ which ultimately rests upon the internalisation of Murray's definition of a 1-gerbe (resp.\ that of a $p$-gerbe) in $\,\sLieGrp$,\ is the first distinctive feature of the geometrisation scheme advocated.

The r\^ole of the above categorification in establishing the descent of $\,\cG_1\,$ to $\,\bT//\G_{\rm KR}\,$ becomes apparent upon realising that --  $\,\G_{\rm KR}\,$ being \emph{discrete} -- the $\,\Phi_t\,$ \emph{with} $\,t\in\G_{\rm KR}\,$ collectively compose a candidate 1-cell $\,\Upsilon\,$ of a $\G_{\rm KR}$-equivariant structure,\ {\it cp.}\ \cite{Runkel:2008gr}.\ At this stage,\ it remains to identify its 2-cell $\,\upsilon$,\ represented,\ in turn,\ by a $\G_{\rm KR}^{\x 2}$-indexed family $\,\upsilon_{\g_1,\g_2}:\la_{\g_1^{-1}}^*\Phi_{\g_2}\circ\Phi_{\g_1}\cong\Phi_{\g_2\cdot\g_1}\,$ of 2-cells in $\,\gt{Grb}_\nabla(\bT)$.\ To this end,\ we employ the bifunctoriality of pullback together with the former identification of $\,\Upsilon\,$ to rewrite the last definition as $\,\upsilon_{\g_1,\g_2}:\id_{\cG_1}\circ\id_{\cG_1}\cong\id_{\cG_1}$,\ whereupon it transpires that the $\,\upsilon_{\g_1,\g_2}\,$ are all given by the natural invertible right-unit 2-cell $\,\rho_\Psi:\id_{\cG_1}\circ\Psi\cong\Psi\,$ in the category $\,\gt{End}(\cG_1)\,$ for $\,\Psi\equiv\id_{\cG_1}\,$ \cite{Waldorf:2007mm}.\ These are determined by $\,\mu_L\equiv\bd1$,\ and hence trivial,\ $\,\upsilon_{\g_1,\g_2}\equiv\bd1$,\ which also implies that they are coherent,\ {\it i.e.},\ $\,\D^{(3)}\upsilon=\id$.\ Thus,\ altogether,\ the triple $\,\cE_{\G_{\rm KR}}\equiv(\cG_1,\{\Phi_\g\equiv\id_{\cG_1}\}_{\g\in\G_{\rm KR}},\{\upsilon_{\g_1,\g_2}\equiv\bd1\}_{\g_1,\g_2\in\G_{\rm KR}})\,$ is a $\G_{\rm KR}$-equivariant structure,\ and $\,\cG_1\,$ acquires the status of a model of a 1-gerbe over $\,\bT//\G_{\rm KR}$.\ Once again,\ the uniqueness of the CaE super-1-gerbe -- it carries the data of its own descent -- hinges on its internalisation in $\,\sLieGrp$.\ In conjunction with the previous observation regarding the categorification of supersymmetry,\ this sets apart the geometrisation scheme adopted and reveals its deeper topological meaning.  

The descent of the $\,\cG_p\,$ to the Rabin--Crane super-orbifold has an important superfield-theoretic implication:\ The triple $\,\cE_{\G_{\rm KR}}\,$ can be used to consistently \emph{define},\ upon pullback to $\,\Si\,$ with $\G_{\rm KR}$-jump defects,\ a super-$\si$-model with iDOFs modelled on $\,\bT//\G_{\rm KR}$,\ even in the absence of a smooth supermanifold structure on the latter,\ {\it cp.}\ the detailed study \cite{Runkel:2008gr}.\ In this manner,\ the geometrisation scheme delineated in this note promises to open a new direction in both:\ the study of non-Rothstein supergeometries \emph{and} the modelling of (super)charged dynamics thereon.

\end{document}